\documentclass[letterpaper,12pt,oneside,amstex]{tesis}

\usepackage[indonesia]{babel}
\usepackage[T1]{fontenc}
\usepackage[latin1]{inputenc}
\usepackage{rotating}
\usepackage{graphicx}
\usepackage{amssymb}
\usepackage{array}
\usepackage{longtable}
\usepackage{setspace}
\usepackage{multind}
\usepackage{textcomp}
\usepackage{ifpdf}
\ifpdf

      \usepackage[pdftex]{hyperref}
\fi
\usepackage{algorithmic}
\usepackage{algorithm}

\makeindex{subject}
\centerchapter
\begin{document}
\nocite{*}
 \normalsize
\pagestyle{empty}
\pagenumbering{roman}
\setcounter{page}{1}
\newpage
\begin{center}
\bfseries
 { \large PERFORMANCE EVALUATION OF PARALLEL MESSAGE PASSING AND THREAD PROGRAMMING MODEL ON MULTICORE ARCHITECTURES}
\vspace{2.5cm}

\normalsize
D.T. Hasta and A.B. Mutiara\\ 



\vspace{1.5cm}

{\small
    GRADUATE PROGRAM,
    GUNADARMA UNIVERSITY, \\ 
    Jl. Margonda Raya No.100, Depok 16424, Indonesia \\
    2010 
}
\end{center}

\pagestyle{plain}
\pagenumbering{roman}
\setcounter{page}{2} 
\newpage 
\addcontentsline{toc}{chapter}{ABSTRACT}
\begin{center}
\begin{large}\textbf{ABSTRACT}\end{large}
\end{center}

\vspace{5mm}

\normalsize
\noindent \textbf{D.T. HASTA, A.B. Mutiara} \\
\textbf{TITLE: PERFORMANCE EVALUATION OF PARALLEL MESSAGE PASSING AND THREAD PROGRAMMING  MODEL ON MULTICORE ARCHITECTURES}

The current trend of multicore architectures on shared memory systems underscores the
need of parallelism. While there are some programming model to express parallelism, thread programming model has become a standard to support these system such as OpenMP, and POSIX threads. MPI (Message Passing Interface) which remains the dominant model used in high-performance computing today faces this challenge.

Previous version of MPI which is MPI-1 has no shared memory concept, and Current MPI version 2 which is MPI-2 has a limited support for shared memory systems. In this research, MPI-2 version of MPI will be compared with OpenMP to see how well does MPI perform on multicore / SMP (Symmetric Multiprocessor) machines.

Comparison between OpenMP for thread programming model and MPI for message passing programming model will be conducted on multicore shared memory machine architectures to see who has a better performance in terms of speed and throughput. Application used to assess the scalability of the evaluated parallel programming solutions is matrix multiplication with customizable matrix dimension.

Many research done on a large scale parallel computing which using high scale benchmark such as NSA Parallel Benchmark (NPB) for their testing standarization [1]. This research will be conducted on a small scale parallel computing that emphasize more on the performance evaluation between MPI and OpenMPI parallel programming model using self created benchmark.\\

\noindent Bibliography (2007-2010)




\addcontentsline{toc}{chapter}{TABLE OF CONTENTS}
\tableofcontents{}

\listoffigures
\addcontentsline{toc}{chapter}{LIST OF FIGURES}

\listoftables
\addcontentsline{toc}{chapter}{LIST OF TABLES}




\newpage
\pagestyle{headings}
\pagenumbering{arabic} 
\setcounter{page}{1} 

\chapter{INTRODUCTION}

\normalsize
\section{Background}

The growth of multicore processors has increased the need for parallel programs on the largest to the smallest of systems (clusters to laptops).There are many ways to express parallelism in a program. In HPC (High Performance Computing), the MPI (Message Passing Interface) has been the main tool for parallel message passing programming model of most programmers \cite{WikiDist:2010}.

A multi-core processor looks the same as a multi-socket single-core server to the operating system. (i.e. before multi-core, dual socket servers provided two processors like today's dual core processors) Programming in this environment is essentially a mater of using POSIX threads. 

Thread programming can be difficult and error prone. OpenMP was developed to give programmers a higher level of abstraction and make thread programming easier. Accordance to multicore trend growth, parallel programming using OpenMP gains popularity between HPC developers. 

Together with the growth of thread programming model on shared memory machines, MPI which has been intended for parallel distributed systems since MPI-1, also has improved to support shared memory systems. The principal MPI-1 model has no shared memory concept, and MPI-2 has only a limited distributed shared memory concept. Nonetheless, MPi programs are regularly run on shared memory computers.

MPI performance for shared memory systems, will be tested on cluster of shared memory machines. OpenMPI will be used as a reference on the same multicore systems with MPI clusters (both MPI and OpenMPI will have an equal amount of core workes).

Application used as a testing is N X N rectangular matrix multiplication with adjustable matrix dimension N ranging from 10 to 2000. For OpenMP, a single multicore machines with two worker cores will be used to calculated the matrix. For the MPI, two multicore machines with three worker cores will be used (one as a master process who decompose the matrix to sub - matrix and distrbute it to two other worker process and compose the final result matrix from the sub - matrix multiplication done by its two worker process).

Parameter result which can be obtained from this test are amount of floating point operation per second (FLOPS) which in this case is matrix multiplication, Program Running Time, and Speedup.  For MPI, two machines used for testing having quite similiar performance (Memory and CPU performance). MPI testing done via LAN cable medium transmission to achieve best run time performance and minimizing time communication between process.

\section{Related Research}

There are already related research topik in this area. One of them did the testing by using certain parallel benchmark such as NAS Parallel benchmark (NPB) to standarize the evaluation on large clusters and ten to hundreds of CPU cores \cite{europvm:2009}, which can produce large speedup and throughput.

This research is simpler compared than the previous research which uses tens or hundreds machine resources and complicated benchmark. This research focused on how does different parallel programming model can affect the program performance. Testing in this research done by using an self created benchmark which count the program running time and matrix multiplication operation per second.

\section{Research Significance}
This research describes how does workshare process done on different parallel programming model. This research gives comparative result between message passing and shared memory programming model in runtime and amount of throughput. Testing methodology also simple and has high usability on the available resources.

\section{Research Problem}

Problem covered in this research are:
\begin{enumerate}
  \item How does different parallel programming model influenced parallel performance on different memory architecture?
  \item How does workshare construct differ between shared and distributed shared memory systems?
\end{enumerate}

\section{Research Purposes}
Objectives of this research are:
\begin{enumerate}
 \item Evaluating parallel performance between thread and message passing programming model.
 \item Evaluating parallel algorithm workshare between threads and message passing programming model.
\end{enumerate}

\section{Scope of the Research}

Testing experiment conducted on a single multicore shared memory machine which consist of two cores for thread programming model (OpenMP). And two identical multicore shared memory machine with two cores on each machine for message passing programming model without expressing thread safety level. Matrix Multiplication Program used for testing also has a limited dimension which is 2000, because of the machine power limitation. Testing parameters generated are amount of floating point operation per second (FLOPS) which in this case is matrix multiplication, Program Running Time, and Speedup 



\chapter{FUNDAMENTAL THEORY}

Parallel computing is a form of computation in which many calculations are carried out simultaneously, operating on the principle that large problems can often be divided into smaller ones, which are then solved concurrently ("in parallel"). Parallel computing is done by a certain amount of parallel computers. Each of parallel computer may has different CPU core and memory architecture

Parallel computers can be roughly classified according to the level at which the hardware supports parallelism. Currently there are three types which are shared memory (which usually has multiple core processor), distributed memory (clusters, MPPs, and grids), and Distributed shared memory (cluster of Shared memory systems).

\section{Shared Memory Systems}

In computer hardware, shared memory refers to a (typically) large block of random access memory that can be accessed by several different central processing units (CPUs) in a multiple-processor computer system.

A shared-memory parallel computer whose individual processors share memory
(and I/O) in such a way that each of them can access any memory location with
the same speed; that is, they have a uniform memory access (UMA) time.

Each of individual processor in shared memory system has a small and fast private cache memory. Cache memory used to supply each core processor with data and instruction at high rates. This is because fetching data done from processor to main memory directly is slower than fetching it from cache memory.

The issue with shared memory systems is that many CPUs need fast access to memory and will likely cache memory \cite{WikiShared:2010}, which has two complications:

\begin{itemize}
\item CPU-to-memory connection becomes a bottleneck. Shared memory computers cannot scale very well. Most of them have ten or fewer processors.
\item Cache coherence: Whenever one cache is updated with information that may be used by other processors, the change needs to be reflected to the other processors, otherwise the different processors will be working with incoherent data. Such coherence protocols can, when they work well, provide extremely high-performance access to shared information between multiple processors. On the other hand they can sometimes become overloaded and become a bottleneck to performance.
\end{itemize}

\begin{figure}[!htbp]
    \begin{center}
      \includegraphics[scale=0.4]{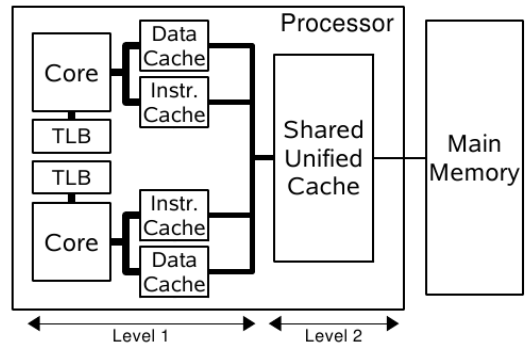}\\
      \caption{Block diagram of a generic, cache-based dual core processor} \label{fig:fig1}
\label{fig:result}
    \end{center}
\end{figure}

However, to avoid memory inconsistency as already mentioned above, there is a cache memory which can be shared to all processor. Shared cache memory can be used for each core processor to write and read data. Figure \ref{fig:fig1} gives information about cache memory.

\section{Distributed Memory}

In computer science, distributed memory refers to a multiple-processor computer system in which each processor has its own private memory. In other words each processor will resides on different computer machine. Computational tasks can only operate on local data, and if remote data is required, the computational task must communicate with one or more remote processors.

\begin{figure}[!htbp]
    \begin{center}
      \includegraphics[scale=0.4]{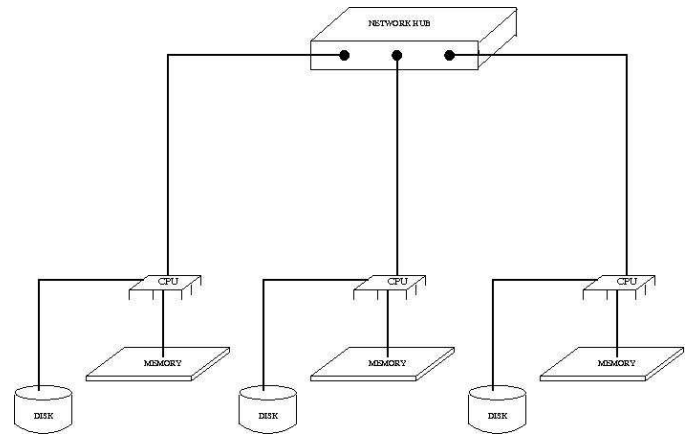}\\
      \caption{An illustration of a distributed memory system of three computers} \label{fig:fig2}
\label{fig:result}
    \end{center}
\end{figure}

In a distributed memory system there is typically a processor, a memory, and some form of interconnection that allows programs on each processor to interact with each other. The interconnect can be organised with point to point links or separate hardware can provide a switching network. The network topology is a key factor in determining how the multi-processor machine scales. The links between nodes can be implemented using some standard network protocol (for example Ethernet), etc. Fig \ref{fig:fig2} shows distributed memory sytems architecture.

In contrast, a shared memory multi processor offers a single memory space used by all processors. Processors do not have to be aware where data resides, except that there may be performance penalties, and that race conditions are to be avoided.

\section{Distributed Shared Memory}

Distributed Shared Memory (DSM), also known as a distributed global address space (DGAS), is a term in computer science that refers to a wide class of software and hardware implementations, in which each node of a cluster has access to shared memory in addition to each node's non-shared private memory.

\begin{figure}[!htbp]
    \begin{center}
      \includegraphics[scale=0.4]{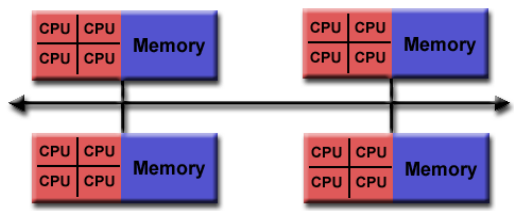}\\
      \caption{An illustration of a distributed shared memory system} \label{fig:figdsm}
\label{fig:result}
    \end{center}
\end{figure}

The shared memory component is usually a cache coherent SMP machine. Processors on a given SMP can address that machine's memory as global. The distributed memory component is the networking of multiple SMPs. SMPs know only about their own memory - not the memory on another SMP. Therefore, network communications are required to move data from one SMP to another. Figure \ref{fig:figdsm} describes about distributed shared memory.

\normalsize
\section{Parallel Programming Models}

All of those parallel hardware classification needs programming language which has a capabilty to share or divided the work among processors. Concurrent programming languages, libraries, APIs, and parallel programming models have been created for programming parallel computers. These can generally be divided into classes based on the assumptions they make about the underlying memory architecture which are shared memory, distributed memory, or shared distributed memory.

A parallel programming model is a set of software technologies to express parallel algorithms and match applications with the underlying parallel systems. It encloses the areas of applications, programming languages, compilers, libraries, communications systems, and parallel I/O.

Parallel models are implemented in several ways: as libraries invoked from traditional sequential languages, as language extensions, or complete new execution models. They are also roughly categorized for two kinds of systems: shared-memory system and distributed-memory system, though the lines between them are largely blurred nowadays.

Shared memory programming languages communicate by manipulating shared memory variables through threads. Threads used as substasks which carry instruction process to one / more core processor. however, one thread only can carry one instructions process at a certain time. In other words, multiple threads can carry multiple instruction process Distributed memory uses message passing. POSIX Threads and OpenMP are two of most widely used shared memory APIs, whereas Message Passing Interface (MPI) is the most widely used message-passing system API.

Shared memory systems use . whereas distributed memory systems use message passing task and communication carried out by message passing over network transmission.

A programming model is usually judged by its expressibility and simplicity, which are by all means conflicting factors. The ultimate goal is to improve productivity of programming.

\subsection{OpenMP}

OpenMP (Open Multi-Processing) is an application programming interface (API) that supports multi-platform shared memory multiprocessing programming in C, C++ and Fortran on many architectures, including Unix and Microsoft Windows platforms. It consists of a set of compiler directives, library routines, and environment variables that influence run-time behavior.

\begin{figure}[!htbp]
    \begin{center}
      \includegraphics[scale=0.6]{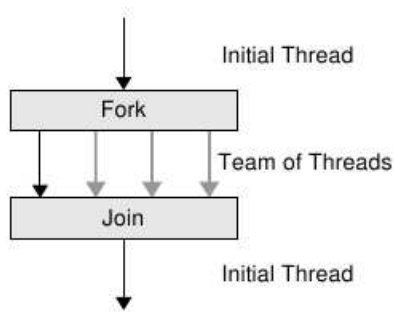}\\
      \caption{The fork-join programming model supported by OpenMP} \label{fig:fig3}
\label{fig:result}
    \end{center}
\end{figure}

OpenMP is an implementation of multithreading, a method of parallelization whereby the master "thread" (a series of instructions executed consecutively) "forks" a specified number of slave "threads" and a task is divided among them. The threads then run concurrently, with the runtime environment allocating threads to different processors. Hence, OpenMP is one of thread based parallel programming which will be used in this research. Figure \ref{fig:fig3} gives a better understanding about multithreading.

OpenMP use pragma directives to express parallelism in the code block. Parts of the program that are not enclosed by a parallel construct will be executed serially. When a thread encounters this construct, a team of threads is created to execute the associated parallel region, which is the code dynamically contained within the parallel construct. But although this construct ensures that computations are performed in parallel, it does not distribute the work of the
region among the threads in a team. In fact, if the programmer does not use the appropriate syntax to specify this action, the work will be replicated. At the end of a parallel region, there is an implied barrier that forces all threads to wait until the work inside the region has been completed. Only the initial thread continues execution after the end of the parallel region.

The thread that encounters the parallel construct becomes the master of the new team. Each thread in the team is assigned a unique thread number (also referred to as the "thread id") to identify it. They range from zero (for master thread) up to one less than the number of threads within the team, and they can be accessed by the programmer.

\subsection{MPI}

Message Passing Interface (MPI) is an API specification that allows computers to communicate with one another. It is used in computer clusters and supercomputers. MPI is a language-independent communications protocol used to program parallel computers. Both point-to-point and collective communication are supported. MPI "is a message-passing application programmer interface, together with protocol and semantic specifications for how its features must behave in any implementation."MPI's goals are high performance, scalability, and portability.

MPI is not sanctioned by any major standards body; nevertheless, it has become a de facto standard for communication among process that model a parallel program running on a distributed memory system. Actual distributed memory supercomputers such as computer clusters often run such programs. The principal MPI-1 model has no shared memory concept, and MPI-2 has only a limited distributed shared memory concept. Nonetheless, MPI programs are regularly run on shared memory computers.

The MPI interface is meant to provide essential virtual topology, synchronization, and communication functionality between a set of processes (that have been mapped to nodes/servers/computer instances) in a language-independent way, with language-specific syntax (bindings), plus a few language-specific features. MPI programs always work with processes, but programmers commonly refer to the processes as processors. Typically, for maximum performance, each CPU (or core in a multi-core machine) will be assigned just a single process. This assignment happens at runtime through the agent that starts the MPI program (i.e. MPI daemon), normally called mpirun or mpiexec. Computer machine which initates MPI ring daemon will has process manager in its core CPU. Process manager identified with ID 0 and all of his worker have ID greater than 0.

The initial implementation of the MPI 1.x standard was MPICH, from Argonne National Laboratory (ANL) and Mississippi State University. ANL has continued developing MPICH for over a decade, and now offers MPICH 2, implementing the MPI-2.1 standard.  

\chapter{WORKSHARE METHODOLOGY}

\normalsize
\section{Matrix Multiplication Workshare Algorithm }
\begin{figure}[!htbp]
    \begin{center}
      \includegraphics[scale=0.4]{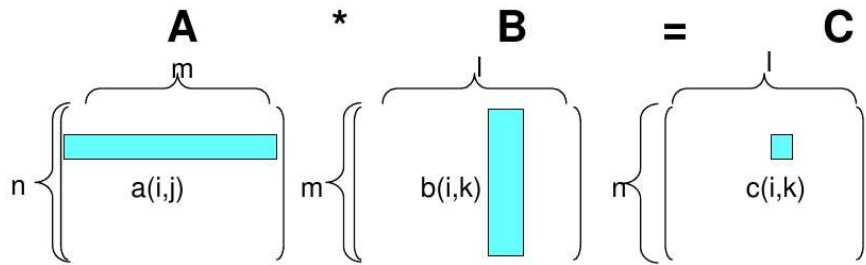}\\
      \caption{Matrix Multiplication Structure} \label{fig:fig4}
\label{fig:result}
    \end{center}
\end{figure}

Matrix multiplication structure is as defined in Figure \ref{fig:fig4}

Multiplying two $N x N$ matrices in sequential algorithm \ref{alg:alg1} takes obviously for each element $N$ multiplications and $N - 1$ additions. Since there are $N^2$ elements in the matrix this yields a total of $N^2 * (2N - 1)$ floating-point operations, or about $2N^3$ for large $N$, that is , ${\bigodot (N^{3})}$.

Parallel algorithm workshare does not change matriks multiplication arithmetic operations. It only change the execution sequence for multiple processors. However, the complexity / operation count will change because of the workshare between core processors.

\begin{algorithm}
\caption{Matrix Multiplication Algorithm}
\label{alg:alg1}
\begin{algorithmic}
\FOR{$i=1$ to $n$}
\FOR{$k=1$ to $l$}
\FOR{$j=1$ to $m$}
\STATE $c(i,k) \leftarrow c(i,k) + a(i,j) \times b(j,k)$
\ENDFOR
\ENDFOR
\ENDFOR
\end{algorithmic}
\end{algorithm}

\subsection{MPI Workshare Algorithm}

MPI shared their work among processing units, by using message passing across network. Process identification between core CPU (both on the same computer and different computer) is similiar with OpenMP (i.e. ID = 0 master process, ID > 0 worker process).

Parallel proramming model using message passing is similiar with unix socket programming in which process manager can send a chunked task to all of his worker process and receive computation result from all of his worker.

Matrix Multiplication between two rectangular matrix can be shared between process using a certain rule. One of the rule is to make the master process as the process manager which divided and distribute matrix elements according to the number of CPU core workers. Thus, if there is 4 process used, 1 will be used as process manager and the other 3 will be used as worker process. Process ID 0 will become process manager and process ID 1 to 2 as workers. There are several MPI model to distribute matrix among the worker processes and One of them is row based distribution.

In row based distribution, For example there are two $N x N$ matrix, $A_i{}_j$ and $B_i{}_j$, which will be multiplicated. All of matrix $B_i{}_j$ elements (rows and collumns) will be sent to all worker processes which is done by master process. For matrix $A_i{}_j$, before it is sent, it will be first divided according to its amount of row. For example if matrix $A_i{}_j$ has 4 rows and there are 3 workers available, each process will has 4/3 which is 1 row with another 1 residual row value generated from the arithmetic division. The residual row value will be added to the worker process which has ID lower or equal than amount of residual row value in a for repetition order (start from worker ID 1 to 2). Thus. 1 residual row value will be added to worker process ID 1.

\begin{figure}[!htbp]
    \begin{center}
      \includegraphics[scale=0.4]{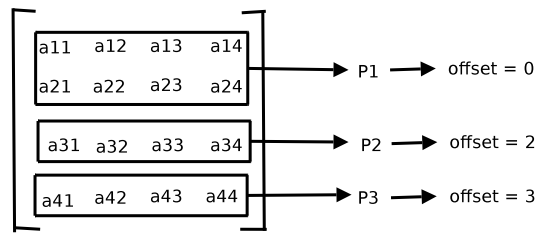}\\
      \caption{Matrix A Row Based Division} \label{fig:fig5}
\label{fig:result}
    \end{center}
\end{figure}

\begin{figure}[!htbp]
    \begin{center}
      \includegraphics[scale=0.4]{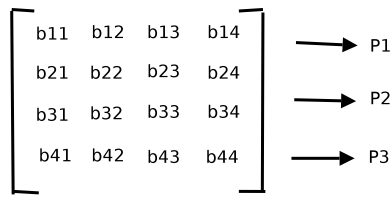}\\
      \caption{Matrix B Distribution} \label{fig:fig6}
\label{fig:result}
    \end{center}
\end{figure}

Figure \ref{fig:fig5} gives row based distribution analogy. Worker process ID 1 work on 2 rows because there is a residual value from the arithmetic division operation. Another reason why worker ID 1 which receives extra row is because worker ID 1 is the first worker process found on the iteration.

Offset variable value will be added for each row sent to worker process.  Note that offset varible has a crucial role for keeping track of matrix A row index so that each worker process knows which row index needs to be worked on. While Figure \ref{fig:fig6} shows how all of rows and collumns of matrix b sent to all worker process.

\begin{figure}[!h]
    \begin{center}
      \includegraphics[scale=0.25]{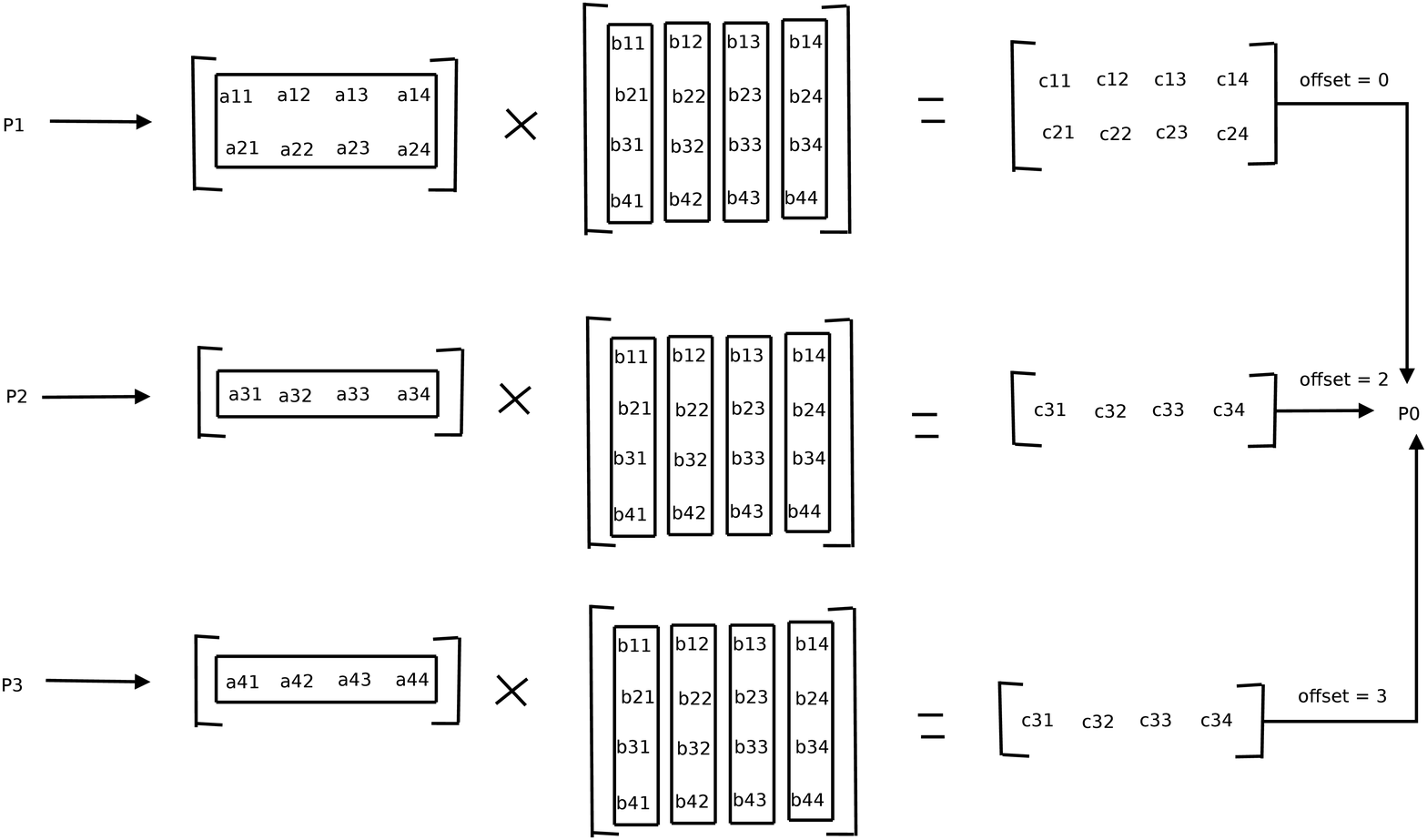}\\
      \caption{Matrix Multiplication For Each Worker Process} \label{fig:fig7}
\label{fig:result}
    \end{center}
\end{figure}

Matrix $C_i{,}_j$ with  $N x N$ dimensions will be used to hold the matrix result elements. The computation of a single $_i$ row element $C_1{,}_j$ (for $_j$ = 1,2,.. N) requires an entire matrix element of $B_i{}_j$ and a subset row element of $A_q{}_j$ (where $_q \subset{} _i$), respectively. Beacuse each worker process has those required element, For number $_P$ process used, each $_P{-}_1$ worker process can compute row element of $C_{(i/(_P-{}_1))+extra}{,}_j$ (for $_j$ = 1,2,.. N) (where $extra$ is an residual variable value and may has different value for different $_P{-}_1$). Matrix computation for each of worker process is shown in Figure \ref{fig:fig7}.

After the sub process multiplication is done in each process, they will send back their matrix result and offset variable value to the master process. For each value received from worker process, master process will generate the final matrix result according to the number offset variable.

The most important thing in this matrix operation division is that there is no dependent data between each sub process matrix multiplication. Thus, there is not any data synchronization needed between sub process, which can reduce program complexity.

\begin{algorithm}
\caption{MPI Master Process Algorithm}
\label{alg:alg2}
\begin{algorithmic}
\begin{scriptsize}
\STATE \COMMENT{Master Process}
\IF{$taskid = 0$}
\STATE \COMMENT{Initializing MatrixA}
\FOR{$i=0$ to $mtrksA\_row - 1$}
\FOR{$j=0$ to $mtrksA\_col - 1$}
\STATE $a[i][j] \leftarrow RAND()$
\ENDFOR
\ENDFOR

\COMMENT{Initializing MatrixB}
\FOR{$i=0$ to $mtrksA\_col - 1$}
\FOR{$j=0$ to $mtrksB\_col - 1$}
\STATE $b[i][j] \leftarrow RAND()$
\ENDFOR
\ENDFOR

\COMMENT{Start Counting Running Time}
\STATE $t\_time \leftarrow MPI\_Wtime()$

\COMMENT{Send matrix data to the worker tasks}
\STATE $averow \leftarrow mtrksA\_row / numworkers$
\STATE $extra \leftarrow mtrksA\_row \% numworkers$
\STATE $offset \leftarrow 0$

\COMMENT{Set the message tag type with value 1, mark the message sent by master process}
\STATE $mtype \leftarrow 1$
\FOR{$dest=1$ to $numworkers$}
\IF{$dest \leq extra$}
\STATE $rows \leftarrow averow + 1$
\ELSE
\STATE $rows \leftarrow averow$
\ENDIF

\COMMENT{Master Sending Process}
\STATE $MPI\_Send(\&offset, 1, MPI\_INT, dest, mtype, MPI\_COMM\_WORLD)$
\STATE $MPI\_Send(\&rows, 1, MPI\_INT, dest, mtype, MPI\_COMM\_WORLD)$
\STATE $count \leftarrow rows * mtrksA\_col$
\STATE $MPI\_Send(\&a[offset][0], count, MPI\_DOUBLE, dest, mtype, MPI\_COMM\_WORLD)$
\STATE $count \leftarrow mtrksA\_col * mtrksB\_col$
\STATE $MPI\_Send(\&b, count, MPI\_DOUBLE, dest, mtype, MPI\_COMM\_WORLD)$

\STATE $offset \leftarrow offset + rows$
\ENDFOR

\COMMENT{Wait for results from all worker tasks}
\STATE $mtype \leftarrow 2$
\FOR{$i=1$ to $numworkers$}
\STATE $source\_prcs \leftarrow i$
\STATE $MPI\_Recv(\&offset, 1, MPI\_INT, source\_prcs, mtype, MPI\_COMM\_WORLD, \&status)$
\STATE $MPI\_Recv(\&rows, 1, MPI\_INT, source\_prcs, mtype, MPI\_COMM\_WORLD, \&status)$
\STATE $count \leftarrow rows * mtrksB\_col$
\STATE $MPI\_Recv(\&c[offset][0], count, MPI\_DOUBLE, source\_prcs, mtype, MPI\_COMM\_WORLD, \&status)$
\ENDFOR

\COMMENT{Stop Counting Running Time}
\STATE $t\_time \leftarrow MPI\_Wtime() - t\_time$

\COMMENT{Counting Amont Of Matrix Multiplication}
\STATE $ops \leftarrow (2 * (pow(NRA, 3)) - (pow(NRA, 2)))$

\COMMENT{Counting Amont Of Matrix Multiplication Per Second}
\STATE $rate \leftarrow ops / t\_time / 1000000.0$
\ENDIF
\end{scriptsize}
\end{algorithmic}
\end{algorithm}

\begin{algorithm}
\caption{MPI Worker Process Alogrithm}
\label{alg:alg3}
\begin{algorithmic}
\begin{scriptsize}
\STATE \COMMENT{Worker Process}
\IF{$taskid > 0$}
\STATE $mtype \leftarrow 1$
\STATE $source\_prcs \leftarrow 0$
\STATE \COMMENT{Retreiving values sent by master process}
\STATE $MPI\_Recv(\&offset, 1, MPI\_INT, source\_prcs, mtype, MPI\_COMM\_WORLD, \&status)$
\STATE $MPI\_Recv(\&rows, 1, MPI\_INT, source\_prcs, mtype, MPI\_COMM\_WORLD, \&status)$
\STATE $count \leftarrow rows * mtrksA\_col$
\STATE $MPI\_Recv(\&a, count, MPI\_DOUBLE, source\_prcs, mtype, MPI\_COMM\_WORLD, \&status)$
\STATE $count \leftarrow mtrksA\_col * mtrksB\_col$
\STATE $MPI_Recv(\&b, count, MPI\_DOUBLE, source\_prcs, mtype, MPI\_COMM\_WORLD, \&status)$
\STATE \COMMENT{Sub matrix multiplication Calculation}
\FOR{$k=0$ to $mtrksB\_col - 1$}
  \FOR{$i=0$ to $rows - 1$}
    \STATE $c[i][k] \leftarrow 0.0$;
    \FOR{$j=0$ to $mtrksA_col - 1$}
      \STATE $c[i][k] \leftarrow c[i][k] + a[i][j] * b[j][k]$
    \ENDFOR
  \ENDFOR
\ENDFOR
\STATE \COMMENT{Sends the sub matrix back to the master process}
\STATE $MPI\_Send(\&offset, 1, MPI\_INT, MASTER, 2, MPI\_COMM\_WORLD)$
\STATE $MPI\_Send(\&rows, 1, MPI\_INT, MASTER, 2, MPI\_COMM\_WORLD)$
\STATE $MPI\_Send(\&c, rows * NCB, MPI\_DOUBLE, MASTER, 2, MPI\_COMM\_WORLD)$
\ENDIF
\end{scriptsize}
\end{algorithmic}
\end{algorithm}

MPI Matrix Multiplications algorithms workshare is divided into two parts, one is for master process shown in Algorithm \ref{alg:alg2} and the other is for work process shown in Algorithm \ref{alg:alg3}.

Master process algorithm steps are:
\begin{enumerate}
\item Value of matrixes is initialized using random function.
\item Wall time calulation process started using MPI\_Wtime() function.
\item Row for each worker process is calculated including addtional residual value.
\item Matrix A sub rows sent to all of worker process according to the row and offset variable. Offset variable will be iterated for each sent process.
\item All of Matrix B elements sent to all of worker process.
\item Master process wait to receive all of sub matrix results which will be sent from all worker processes.
\item Wall time calulation process stopped. Time interval between end and start time is calculated.
\item Total matrix operation is calculated using formula $N^2 * (2N - 1)$ in floating point type variable
\item Matrix operation per second in FLOPS is calculated by dividing the total matrix operations by the matrix runtime interval. For simplicity, FLOPS is converted into MFLOPS (i.e. Mega Floating Point Operation Per Second) by dividing it again with ($10^6$).
\end{enumerate}

Worker process algorithm steps are:
\begin{enumerate}
\item Each worker process receive a subset rows of Matrix A, according to the offset variable sent from master process.
\item Each worker process receive all elements (rows * collumns) of Matrix B which is sent from master process.
\item Matrix multiplication process is done for each worker process.
\item Each worker process send their sub rows of matrix C back to the master process.
\end{enumerate}

In DSM machine architectures, communication cost is classified into two different types. First is communication cost between process which located on different machines, and second is communication cost between process which located on same machines.

For Commmunication cost between worker and master process which located on different machines over the network for matrix distributions, can be roughly calculated as:
\begin{itemize}
\item Cost distributing sub matrix A and matrix B to all worker process: $(_P{}_-{}_1)\ *\ ((N^2) \ +\ (N/(_P{}_-{}_1)))\ *\ _t{}_c$. Which is equal to $({}(_pN^2)\ -\ (N^2)\ +\ N)\ *\ _t{}_c$ (Where $_t{}_c$ represents the time it takes to communicate one datum between processors over the network)
\item Cost for receiving matrix result from all worker process: $(_P{}_-{}_1)\ *\ (N/(_P{}_-{}_1))\ *\ _t{}_c \rightarrow N\ *\ _t{}_c$.
\item Total Communication cost: $({}(_pN^2)\ -\ (N^2)\ +\ 2N)\ *\ _t{}_c$.
\end{itemize}

For Commmunication cost between worker and master process which located on the same machines, its distribution process steps can be assumed to be same with the communcation over the network with execption that time takes to communicate one datum between processors is not $_t{}_c$ but $_t{}_f$. Where $_t{}_f$ is represents the time it takes to communicate one datum between processors over the shared memory which is faster than $_t{}_c$. Thus communication over the shared memory can be calculated as $({}(_pN^2)\ -\ (N^2)\ +\ 2N)\ *\ _t{}_f$.

Total communication cost between worker and master which located over the network and shared memory can be calculated as $({}(_pN^2)\ -\ (N^2)\ +\ 2N)\ *\ _t{}_c\ +\ ({}(_pN^2)\ -\ (N^2)\ +\ 2N)\ *\ _t{}_f$

Matrix multiplication complexity in this parallel program is divided into amount of worker process which is $_P{}_-{}_1$ from the total amount of $_P$ process used. Thus, total matrix multiplication complexity in MPI for each worker process can be defined as ${\bigodot (N^{3} / _P{}_-{}_1)}$.

\subsection{OpenMP Workshare Algorithm}

Because OpenMP is an implementation of multithreading which uses multiple thread as it instruction carrier, OpenMP share their work among the amount of threads used in a parallel region. Thread classifed into two types: master thread and worker thread.

By default, each thread executes the parallelized section of code independently. "Work-sharing constructs" can be used to divide a task among the threads so that each thread executes its allocated part of the code. Both Task parallelism and Data parallelism can be achieved using OpenMP in this way.

To determine how worksharing is done in OpenMP, OpenMP offer worksharing construct feature. Using worksharing construct, programmer can easily distribute work to each thread in parallel region in a well ordered manner. Currently OpenMP support four worksharing construct.

\begin{itemize}
\item omp for: used to split up loop iterations among the threads, also called loop constructs.
\item sections: assigning consecutive but independent code blocks to different threads
\item single: specifying a code block that is executed by only one thread, a barrier is implied in the end
\item master: similar to single, but the code block will be executed by the master thread only and no barrier implied in the end.
\end{itemize}

Since OpenMP is a shared memory programming model, most variables in OpenMP code are visible to all threads by default. But sometimes private variables are necessary to avoid race conditions and there is a need to pass values between the sequential part and the parallel region (the code block executed in parallel), so data environment management is introduced as data sharing attribute clauses by appending them to the OpenMP directive. The different types of clauses are:

\begin{itemize}
\item shared: the data within a parallel region is shared, which means visible and accessible by all threads simultaneously. By default, all variables in the work sharing region are shared except the loop iteration counter.
\item private: the data within a parallel region is private to each thread, which means each thread will have a local copy and use it as a temporary variable. A private variable is not initialized and the value is not maintained for use outside the parallel region. By default, the loop iteration counters in the OpenMP loop constructs are private.
\item default: allows the programmer to state that the default data scoping within a parallel region will be either shared, or none for C/C++, or shared, firstprivate, private, or none for Fortran. The none option forces the programmer to declare each variable in the parallel region using the data sharing attribute clauses.
\item firstprivate:like private except initialized to original value.
\item lastprivate: like private except original value is updated after construct.
\item reduction: a safe way of joining work from all threads after construct.
\end{itemize}

Matrix multiplication workshare between threads in OpenMP, is done for each matrix row similiar to MPI. The difference is that MPI distribute its matrix element by sending it to all worker process, while OpenMP only need to declare the scope of matrix element variable as shared or private.Take an example, matrix multiplication between two $N x N$ matrixes $A_i{}_j$ and $B_i{}_j$ which result will be contained in matrix $C_i{}_j$. Each Matrix has 4 rows ($_i=4$) and number threads used is 2 ($_t=2$).

Rows distribution process done using workshare construct "pragma omp for" which is placed on the outer most of for loop repetitions.  Thus, each thread ($_t$) will responsible for calculating each matrix C $_i$ row for all $_j$ collumn elements.

\begin{figure}[!htbp]
    \begin{center}
      \includegraphics[scale=0.3]{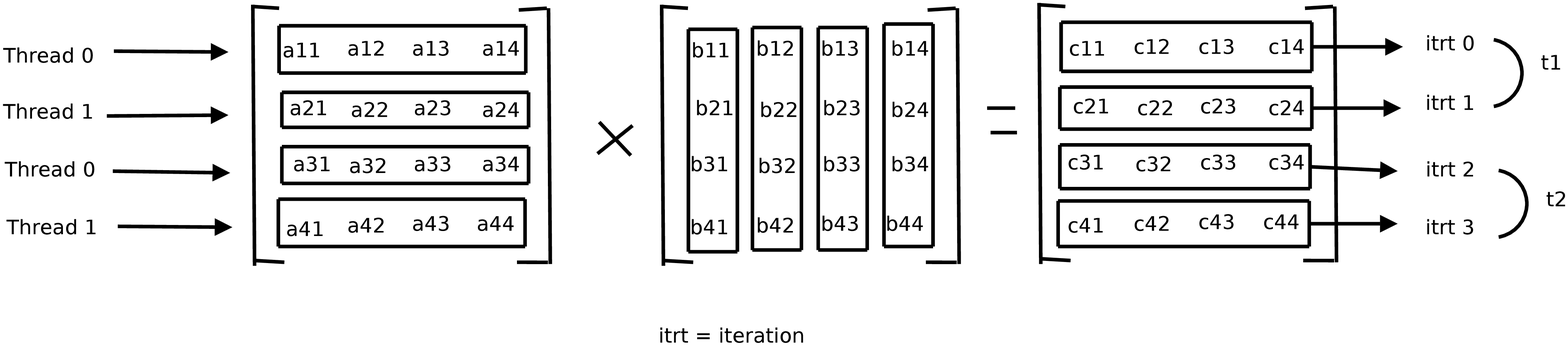}\\
      \caption{OpenMP Matrix Multiplication Algorithm Scheme} \label{fig:fig8}
\label{fig:result}
    \end{center}
\end{figure}

The amount of rows distributed to number of threads is determined by schedule clause. There are 3 types schedule clause (static, dynamic, and guided). Static schedule distribute iterations equally among all of threads (if there is a residual iteratons, threads which has done its job first will be assigned to work on that iterations). Dynamic and guided allows iterations to be assigned to threads according their chunk size defined by programmer. In this program, iteration distribution among threads will be done using static schedule. To understand better how does matrix multiplication done in OpenMP look at Figure \ref{fig:fig8}

Figure \ref{fig:fig8} scheme gives us matrix multiplication process for two different time executions ($t_1$ and $t_2$). because there are 4 rows in matrixes and only 2 thread used in programs, there will be 2 residual rows which will be assigned again to those two threads in different time. in $t_1$, 2 threads (ID 0 and 1) will calculate first and second rows. After those two thread finished (assumed that time execution for each thread is same), in $t_2$ time, those 2 threads will calculate again for third and fourth rows.

Determining how many threads should be used in a parallel region is quite tricky. For the same operation performed by all threads (e.g. matrix multiplication) the most optimal number threads used is the same amount as the total number of cores available. But if in a parallel region consist a various operation (e.g. print, open file, read file, etc) using more than amount of CPU core might be a good idea. In this case, amount of thread can be determined by first, calculating operation cost for each different operations. Thus, number of threads used in matrix multiplication program must equal to the number of CPU cores.

\begin{algorithm}
\caption{OpenMPI Process Alogrithm}
\label{alg:alg4}
\begin{algorithmic}
\STATE \COMMENT{Initalizing Matrix A}
\FOR{$I=0$ to $DIM-1$}	
	\FOR{$J=0$; $DIM-1$}
		\STATE $MatrikA[I][J] \leftarrow RAND()$
	\ENDFOR
\ENDFOR

\COMMENT{Initializing Matrix B}
\FOR{$J=0$ to $DIM-1$}
	\FOR{$K=0$ to $DIM-1$}
		\STATE $MatrikB[J][K] \leftarrow RAND()$
	\ENDFOR
\ENDFOR

\COMMENT{Start Calculating Running Time}
\STATE $tot\_time \leftarrow omp\_get\_wtime()$

\COMMENT{Matrix Multiplication Process}
\STATE \# $pragma\ omp\ parallel\ shared ( MatrikA, MatrikB, MatrikC, DIM )\ private ( I, K, J)$

\# $pragma\ omp\ for$

\FOR{$I=0$ to $DIM-1$}
	\FOR{$K=0$ to $DIM-1$}
		\STATE $MatrikC[I][K] \leftarrow 0$
			\FOR{$J=0$ to $DIM-1$}
				\STATE $MatrikC[I][K] \leftarrow MatrikC[I][K] + MatrikA[I][J] * MatrikB[J][K]$
			\ENDFOR
	\ENDFOR
\ENDFOR

\COMMENT{Stop calculating running time}
\STATE $tot\_time \leftarrow omp\_get\_wtime() - tot\_time$

\COMMENT{Calculate the matrix operation}
\STATE $ops \leftarrow (2 * (pow(DIM, 3)) - (pow(DIM, 2)))$

\COMMENT{Calculate the matrix operation per second}
\STATE $rate \leftarrow ops / tot\_time / 1000000.0$
\end{algorithmic}
\end{algorithm}

The difference between MPI and OpenMP in process management is that MPI needs its master process to do only a specific job, which is distributing  matrix elements, receiving the result calculation, and generating matrix result apart from calculating sub matrix like its workers. The reason behind this is to reduce parallel overhead when calculating a large of matrix dimension over network transmision. Hence, master process can focus only on managing and distributing data.

Unlike MPI in OpenMP, process management and synchronization is done in the same memory (i.e shared memory) and not only master thread but all of thread also responsible of thread synchronization. Hence, master thread can also participate in the matrix multiplication process.

OpenMP matrix multiplication algorithm steps are:
\begin{enumerate}
\item Initializing matrixes value using random function
\item Wall time calulation process started using omp\_get\_wtime() function.
\item Matrix multiplication process for each thread is conducted using pragma om parallel directives and pragma omp for workshare construct.
\item Wall time calulation process stopped. Time interval between end and start runtime calculation process is calculated.
\item Total matrix operation is calculated using formula $N^2 * (2N - 1)$ in floating point type variable
\item Matrix operation per second is calculated in FLOPS and then converted in MFLOPS which same with MPI.
\end{enumerate}

Unlike MPI, which communication is done using message passing, communication cost in OpenMP is conducted between threads which is assumed to be fast and insignificant to the peformance. Thus its time calculation can be ignored in this algorithm.

Matrix multiplication complexity in OpenMP parallel is divided into amount of threads which is $_t{}$ including master threads. Thus, total matrix multiplication complexity in OpenMPI for each thread can be defined as ${\bigodot (N^{3} / _t)}$ (where number of threads is equal to number of CPU cores).  

\chapter{RESULT AND DISCUSSION}

\normalsize
\section{Performance Test}
Matrix multiplication alogrithm tested ranging from dimension 100 up to 2000 on a three different scenario: sequential algorithm, MPI algorithm, and OpenMP algorithm. For OpenMP and sequential program, test was done on a single Intel core duo 1.7 Ghz T5300 laptop with 1GB RAM running on linux SUSE. For MPI program test was done on a two Intel core duo laptops one with frequency 1.7 GHz T5300 laptop with 1GB RAM running on Windows Xp SP3 and another one with frequency 1.83 GHz T5500 with 1GB RAM running on Windows Xp SP3.

Number of threads used in OpenMP is two (one master thread, and the other one is worker thread). unlike OpenMP, number of process used is three (two worker process and one is master process). The reason is already discussed in the previous section. However, the number of worker process / threads which performed the matrix multiplication process is equal for both programming models (i.e. two workers).

Because there are three processes used in MPI which will be distributed on two multicore / SMP machines (i.e each machines will have two cores), one of the two machines will have its both of core CPUs occupied (i.e master process and worker process). Computer which initiate the MPI ring daemon, has a master process in one of its core. Thus, computer machine with master process in it will also has a worker process, and the other machine will only has one worker process.

Stastical analysis conducted on one independent variable (i.e. Matrix Dimension) towards three dependent variable (e.g. runtime program, throughput program, and speedup). Using this statistical analysis, matrix dimension (i.e. as an independent variable) influence towards all of three dependent variable can be seen clearly.

\section{Statistical Analysis Result}

\subsection{Parallel Runtime Towards Matrix Dimension}
\begin{figure}[!h]
    \begin{center}
      \includegraphics[scale=0.6]{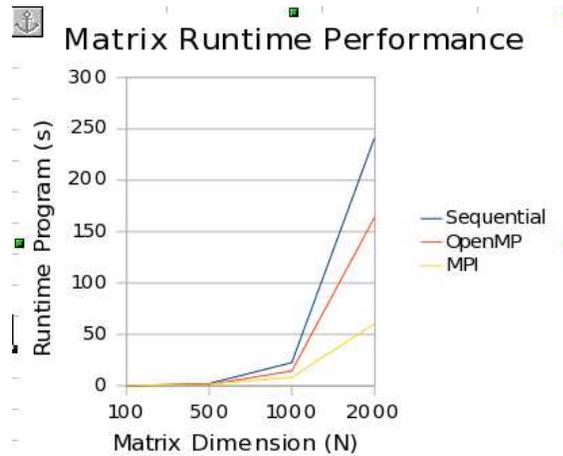}\\
      \caption{Running time comparison with various matrix dimension} \label{fig:fig9}
\label{fig:result}
    \end{center}
\end{figure}

Fig \ref{fig:fig9} gives run time program obtained using wall time function for three different program(e.g. sequential, OpenMP, MPI). Wall time is the actual time taken by a computer to complete a task (i.e matrix multiplication). It is the sum of three terms: CPU time, I/O time, and the communication channel delay (e.g. if data are scattered on multiple machines (MPI)).

In OpenMP, wall time is calculated using omp\_get\_wtime() which starts from when the inital thread enter the parallel region until it exits the parallel region. Thus, process time calculated are thread creation, synchronization and multiplication tasks.

In MPI, wall time is calculated using MPI\_Wtime() which starts from when the master process distribute the work among the worker processes until it receives matrix results sent from all worker processes. Process time calculated in MPI are communication between master - worker process and matrix multiplication for all worker process.

For N = 100, runtime MPI is much slower up to 10 times compared to sequential and OpenMP. However For N = 100 to 2000 MPI runtime is gradually become faster compared to those two. MPI has the fastest runtime performance for N > 500.

\begin{table}[!h]
\centering
\caption{Running time comparison with various matrix dimension} \vspace{.2cm}
\label{tbl:run}
	\begin{tabular}{|l|l|l|l|}
	\hline
	Matrix Dimension (N) & Sequential (s) & OpenMP (s) & MPI (s)\tabularnewline
	\hline
	\hline
	100 & 0.03 & 0.02 & 0.33\tabularnewline
	\hline
	500 & 2.09 & 1.11 &	1.52\tabularnewline
	\hline
	1000 & 22.52 & 14.36 & 8.19\tabularnewline
	\hline
	2000 & 240.97 & 163.6 &	60.19\tabularnewline
	\hline
	\end{tabular}
	\label{tab:run_time}
\end{table}

\subsection{Parallel Throughput Towards Matrix Dimension}
\begin{figure}[!h]
    \begin{center}
      \includegraphics[scale=0.6]{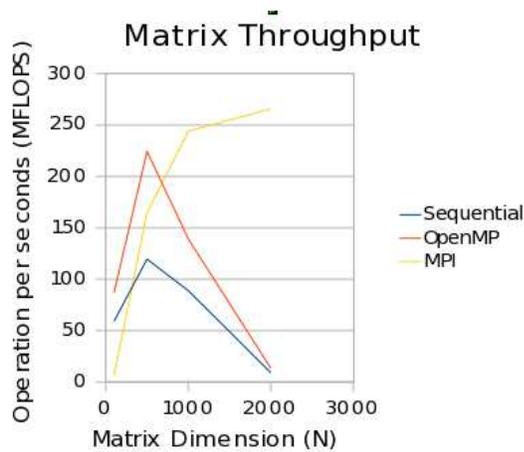}\\
      \caption{Throughput comparison with various matrix dimension} \label{fig:fig10}
\label{fig:result}
    \end{center}
\end{figure}

Fig \ref{fig:fig10} gives througput result for three different program(e.g. sequential, OpenMP, MPI). Throughput in MFLOPS is calculated by dividing number of matrix operations by wall time and $10^6$.

Both sequential and OpenMP has a throughput increase from N = 100 to 500, however, it starts to decrease from N = 500 to 2000. Nevertheless, MPI is different from the other two, It has a continously increased throughput start from N = 100 to 2000. Eventhough, the througput increase from N = 1000 to 2000 is not as significant as before.

\begin{table}[!h]
\centering
\caption{Throughput comparison with various matrix dimension} \vspace{.2cm}
\label{tbl:throu}
	\begin{tabular}{|l|l|l|l|}
	\hline
	Matrix Dimension (N) & Sequential (MFLOPS) & OpenMP (MFLOPS) & MPI (MFLOPS)\tabularnewline
	\hline
	\hline
	100 & 59.08	& 86.82 & 6.07\tabularnewline
	\hline
	500 & 119.54 & 224.35 & 164.85\tabularnewline
	\hline
	1000 & 88.77 & 139.17 & 244.17\tabularnewline
	\hline
	2000 & 8.91	& 13.17 & 265.77\tabularnewline
	\hline
	\end{tabular}
	\label{tab:run_time}
\end{table}

\subsection{Parallel Speedup Towards Matrix Dimension}
\begin{figure}[!h]
    \begin{center}
      \includegraphics[scale=0.6]{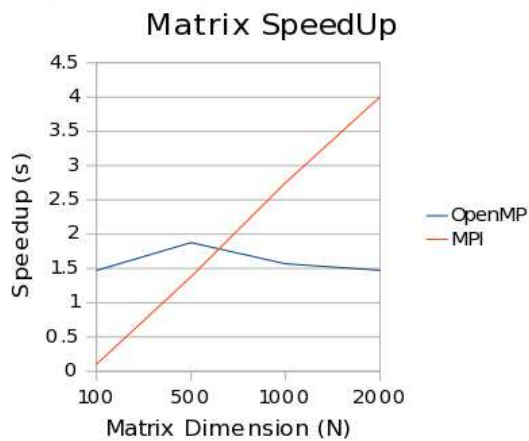}\\
      \caption{Speed comparison with various matrix dimension} \label{fig:fig11}
\label{fig:result}
    \end{center}
\end{figure}

Fig \ref{fig:fig11} gives speedup performance for two program(e.g. OpenMP, MPI) towards sequential program. Speedup performance in this research can be obtained by dividing $wall-clock\ time\ of\ serial\ execution$ with  $wall-clock\ time\ of\ parallel\ execution$

OpenMP has a steady speedup for N = 100 to 2000 which has the average value at 1.5s, while MPI gives a linear speed up growth for N = 500 to 2000 ranging from 1.38s to 4s.
MPI gives no speedup for N = 100 because the matrix calculation is to small compared to the MPI running time and communication time.

\begin{table}[!h]
\centering
\caption{Speed comparison with various matrix dimension} \vspace{.2cm}
\label{tbl:throu}
	\begin{tabular}{|l|l|l|l|}
	\hline
	Matrix Dimension (N) & OpenMP (s) & MPI (s)\tabularnewline
	\hline
	\hline
	100 & 1.47 & 0.1\tabularnewline
	\hline
	500 & 1.88 & 1.38\tabularnewline
	\hline
	1000 & 1.57	& 2.75\tabularnewline
	\hline
	2000 & 1.47	& 4\tabularnewline
	\hline
	\end{tabular}
	\label{tab:run_time}
\end{table}

\chapter{CONCLUDING REMARKS}

\normalsize
\section{Conclusion}
OpenMP workshare between threads in matriks multiplication algorithm, done by using OpenMP FOR workshare construct. OpenMP FOR workshare in matrix multiplication algorithm is placed in the most outer loop of matrix multiplication operation. Using this placement, each OpenMP thread is assigned to work on each matrix C row for all collumns.

MPI workshare in matriks multiplication algorithm done by using send and receive command. Matrix A row will be divided and sent together with all matrix B elements according to the number of worker process both on the different machines and the same machines. Each worker process will work on one row of Matrix A multiplied by all row matrix B elements. If there are  residual row, it will be added one each from the smallest worker process ID to the biggest worker ID.

The performance between OpenMP and MPI programming model is vary for matrix dimension N from 1 to 2000, although many standarizations made for both of parallel programming models (e.g. number of matrix workers, matrix algorithm steps, and machine specifications). Matrix multiplication complexity is divided for the same number of worker (i.e threads if its OpenMP with the complexity of ${\bigodot (N^{3} / _P{}_-{}_1)}$ and process if its MPI with the complexity of  ${\bigodot (N^{3} / _t)}$). Machine specifications used in MPI also comparable with OpenMP which are: Intel Core Duo 1.7 GHz (for OpenMP) and Intel Core Duo 1.7 GHz together with Intel Core Duo 1.83 GHz both with 1 GB of RAM (for MPI).

Performance decline is common in every program testing performance ecspecially when the data testing becomes large. This is due to the resources limitation (CPUs, memory, etc). However for different programming models which use the same resources and algorithm program, there are more reasons than just resources limitations.

For sequential because the worker process only one, its obvious that its overall performance is lower than the other two. For the MPI and OpenMPI, differences can be caused by how fast the workshare is done between worker processes / thread. MPI use message passing in sharing its work across processes which has network communication time over the network medium. In the other hand, OpenMP use thread in sharing its work inside shared memory machine which has no network  communication time. Hence, OpenMP workshare should be faster than MPI. 

This can be seen at Figures \ref{fig:fig9}, \ref{fig:fig10}, and \ref{fig:fig10}. OpenMP gains fast speedup and large throughput for N = 100 to 500 while MPI gains slower but steady speedup and throughput. However, when as N grows larger (i.e N = 500 to 2000) OpenMP performance is gradually become slower while MPI can still keep up with the growth of N.

Besides, the speed of the workshare, performance differences between OpenMP and MPI for large N computation can be caused by core CPUs access to memory. In parallel computing, memory is scalable with number of processors. Thus, increase in the number of processors and the size of memory will also increases the performance scalability. 

MPI distribute its work by copying it on each worker process whether its located on the same memory or different memory machine. Thus, each processor which located on different machine can rapidly access its own memory without interference and without the overhead incurred with trying to maintain cache coherency (i.e MPI provides strong memory locality).

For this research, MPI is tested on distributed shared memory architectures using two SMP (Symmetric Multiprocessor) machines. MPI share two worker process between two different machines, thus MPI distribute the copy data located in different machines. Two worker process can access its data on its own memory which will reduced the overhead half compared on single memory.

OpenMP in the other hand, use a single memory which is shared between core CPU in computer machine (UMA). Hence, the primary disadvantage is the lack of scalability between memory and CPUs. Adding more CPUs can geometrically increases traffic on the shared memory-CPU path, which leads to difficulty maintaining cache coherent systems between processors.

Therefore, the larger memory used in OpenMP, the more congested traffic on the shared memory-CPU path which result in bottleneck. Increase in traffic associated with cache/memory management will produce more parallel overhead while maintaining cache coherency. OpenMP matrix program experienced this problem for N = 500 to 2000. The reason why the performance in OpenMP is decreasing starting from N = 500 to 2000 is because traffic on the shared memory-CPU path is gardually become more congested for memory equal to 1GB. This can be seen at Figure \ref{fig:fig11}, where OpenMP does not give any more speedup than 1.5s.

\section{Future Work}

Message Passing and Thread has been combined in a DSM (Distributed Shared Memory) parallel architecture to achieve a better performance results in nowdays. In this research MPI parallel expression used on shared memory architectures, has not exploited the thread safety  programming explicitly. Using thread safety expression, MPI can explicitly control the threads which running multiple cores accross SMP machines. This Message Passing - thread model also referred as Hybrid parallel programming model. In the next research, Hybrid parallel programming model will be used as evaluation material on DSM (Distributed Shared Memory) architecture.  



\newpage
\addcontentsline{toc}{chapter}{BIBLIOGRAPHY} 
\bibliographystyle{IEEEtranS}
\bibliography{MainTemplateTesis} 


\newpage
\pagestyle{empty}
\include{lampiran}

\end{document}